\documentclass[prl,aps,showpacs,twocolumn,floatfix]{revtex4}

\usepackage{graphicx}
\usepackage[ansinew]{inputenc}
\usepackage{array}
\usepackage{color}
\usepackage{amsmath}
\usepackage{amsxtra}
\usepackage{amstext}
\usepackage{amssymb}
\usepackage{latexsym}
\usepackage{dsfont}
\begin{document}

\title{Multiple membrane cavity optomechanics}

\author{M. Bhattacharya and P. Meystre}
\affiliation{B2 Institute, Department of Physics and College of Optical
Sciences\\The University of Arizona, Tucson, Arizona 85721}

\date{\today}

\begin{abstract}
We investigate theoretically the extension of cavity 
optomechanics to multiple membrane systems. We describe 
such a system in terms of the coupling of the collective 
normal modes of the membrane array to the light fields. 
We show these modes can be optically addressed individually 
and be cooled, trapped and characterized, e.g. via 
quantum nondemolition measurements. Analogies between 
this system and a linear chain of trapped ions or dipolar 
molecules imply the possibility of related applications 
in the quantum regime.
\end{abstract}


\maketitle

Cavity optomechanics is an emerging field at the boundary 
between quantum optics and nanoscience. Resulting in part 
from experimental innovations at the mesoscopic scale, 
optomechanical systems -- mechanical systems that 
can be manipulated by light -- have recently generated
much experimental and theoretical interest 
\cite{cohadon1999,gigan2006,kleckner2006,arcizet2006,schliesser2006,
corbitt2007}. They offer the prospect of realizing quantum
effects at a macroscopic scale \cite{marshall2003}, of 
supplying novel quantum sensors for applications ranging 
from single molecule detection \cite{roukes2006} 
to gravitational wave interferometry 
\cite{courty2003,corbitt2007}, for the quantum control of 
atomic, molecular and optical systems \cite{hansch2007}, 
and for possible new quantum information processing devices 
\cite{mancini2003}.

As cavity optomechanics begins to mature as a field, we
recognize that scalability is an important aspect of any 
technology. In particular, it is immediately relevant to 
possible uses in information processing. This has been 
acknowledged in proposals for constructing quantum computers 
using trapped ions, cavity quantum electrodynamics, 
neutral-atomic lattices, nuclear magnetic resonance, 
spintronics, dipolar molecules, etc., see 
Ref.~\cite{Braunsteinbook2} and references therein. It
is therefore important to investigate the scaling of 
current cavity-based techniques to a larger 
number of optomechanical elements.

\begin{figure}[ht!]
\includegraphics[width=0.44 \textwidth]{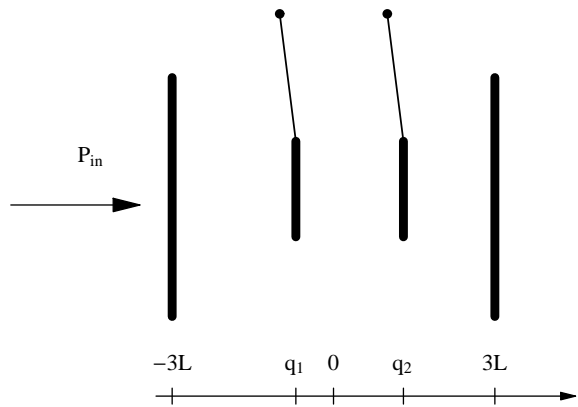}
\caption{\label{fig:cavitypic} A high finesse optical 
cavity with two mirrors fixed at $\pm 3L$, and two 
dielectric membranes centered at $q_{1}\sim -L$ and 
$q_{2}\sim L$ respectively. $P_{in}$ signifies the 
power in the laser light that may be used to drive 
the cavity.}
\end{figure}

In this Letter we consider the optomechanical cooling 
and trapping of a small array of partially transparent 
dielectric membranes inside a high-finesse cavity driven 
by laser radiation
(Fig.~\ref{fig:cavitypic}).
Presenting first the case of two moving membranes, $N=2$, 
where explicit analytical results are readily obtainable,
we show that it is described most conveniently in terms 
of the optical cooling, trapping and measurement of the 
{\em normal modes} of the linear chain formed by the 
membranes. This demonstration depends crucially on the
fact that the optical spectrum of the cavity is symmetric 
in the symmetric normal modes of the membrane array, while 
it does not have well-defined symmetries with respect
to the motion of the individual membranes. We then 
extrapolate these considerations to the case of $N$ 
membranes in a cavity, and draw an analogy to chains 
of trapped ions \cite{zoller1995,wineland1998} and of 
dipolar molecules \cite{demille2002}. We conjecture 
that this analogy may have implications for information 
processing if, as expected, the membranes can be 
experimentally placed in the quantum regime.

Our immediate motivation derives from a pioneering experiment that
demonstrated that a single membrane can be optomechanically cooled
when placed inside a high finesse optical resonator
\cite{thompson2008}. This is the first experiment to show that the
two technologically challenging requirements of high optical and
mechanical quality can be allocated separately to the end-mirrors
and the dielectric membrane respectively, distinct from the
standard two-mirror cavity configuration \cite{cohadon1999}. This
work also pointed out that in addition to the usual linear
coupling of radiation to the displacement $q$ of the membrane that
facilitates back-action cooling, a coupling proportional to $q^2$
is also possible. If the membrane oscillation frequency is larger
than the optical linewidth of the cavity
\cite{kippenberg2007,girvin2007,anetsberger2007}, this quadratic
interaction enables quantum non-demolition measurements of the
energy of the vibrating membrane. Linear couplings do not allow
such measurements \cite{braginsky1980}.

The present work draws on our previous analysis of the case of a
single membrane \cite{mishpm2}, which yields an approximate
microscopic Hamiltonian that describes the optomechanics of the
system. We show that this formalism can in principle be
generalized to $N$ membranes, with the generic features extracted
simply from considering the case $N=2$ (Fig.~\ref{fig:cavitypic}).
More detailed calculations will be presented elsewhere
\cite{mishpm7}.

Our starting point is a cavity with two fixed and perfectly
reflecting end mirrors and two identical vibrating dielectric
membranes, each of reflectivity $R$, mass $m$ and mechanical
frequency $\omega_m$, see Fig.~\ref{fig:cavitypic}. We model the
system with the quantum mechanical Hamiltonian \cite{mishpm2}
\begin{equation}
 \label{eq:Ham}
H= \displaystyle \sum_{j}^{2}\left(\frac{p_{j}^{2}}{2m}+
\frac{1}{2}m \omega_{m}^{2}q_{j}^{2}\right)+
\displaystyle \sum_{i}^{3}\hbar
\omega_{i}(q_{j})a_{i}^{\dagger}a_{i},
\end{equation}
where $p_{j}$ and $q_{j}$, $j=1,2$ denote the momentum and
position of the membranes and $\omega_{i}$, $i=1, 2, 3,$ are the
resonance frequencies of three near-degenerate optical modes with
creation and destruction operators $a_{i}$ and $a_{i}^{\dagger}$
respectively. The commutation rules obeyed by these operators are
$[q_{j},p_{l}]=i\hbar \delta_{jl}$ and $[a_{i},a_{n}^{\dagger}]=
\delta_{in}.$ We discuss shortly why three optical modes are
sufficient for a minimal description of the system.

The first term in Eq.~(\ref{eq:Ham}) is the mechanical energy of
the oscillating membranes, and the second term the energy of
optical modes of the full resonator of length $6L$. The dependence
of the cavity mode frequencies $\omega_{i}$ on the positions
$q_{j}$ of the membranes is central to the description of the
system, since it determines the optomechanical couplings
\cite{mishpm2}. The $\omega_{i}$ are obtained by solving the
classical Maxwell equations for the full resonator. We assume that
the moving membranes are much thinner than an optical wavelength
and model them by spatial delta functions \cite{mishpm2}. For
$R=1$, the resonator consists simply of three uncoupled cavities
whose eigenfrequencies in the absence of membrane motion
$(q_{1,2}=\mp L)$ are threefold degenerate and are given by
\begin{equation}
\label{eq:w}
 \omega_{n}= \frac{n \pi c}{2L}.
\end{equation}
Here $n$ is a positive integer, $c$ is the velocity of light, and
$2L$ is the length of each sub-cavity. This is the reason we
included only three modes in Eq.~(\ref{eq:Ham}); the spectrum may
generally be grouped in such triplets.

For $R \neq 1$, the three resonators are coupled, this coupling
lifting the degeneracy of the frequencies $\omega_n.$ Solving
Maxwell's equations with the appropriate boundary conditions at
the membranes, and neglecting light absorption in these membranes,
we find that the allowed wave-numbers of the complete resonator
can be expressed in terms of the relative coordinate $q=q_1-q_2$
and the center-of-mass (COM) coordinate $Q=(q_{1}+q_{2})/2$ of the
membranes via the trigonometric equation
\begin{eqnarray}
\label{eq:trans}
\sin 2(\theta+3kL)+\sin^{2}\theta
\sin 2k\left (3L-q\right)\nonumber \\
=2 \sin \theta \cos\left(\theta+k q\right)\cos 2kQ,
\end{eqnarray}
where $\sin\theta=\sqrt{R}$.

Equation~(\ref{eq:trans}) is invariant under the transformation $Q
\rightarrow -Q$. Physically this means that the frequency spectrum
is symmetric in the COM motion. We show below that this symmetry
can be exploited to obtain a description of the system purely in
terms of the `phonon' modes associated with the relative and COM
coordinates $q$ and $Q$ separately, that is, excluding any effects
resulting from the coupling between these coordinates. This is in
contrast to an equivalent description in terms of the individual
oscillator coordinates $q_1$ and $q_2$, which typically retains
coupling terms.

\begin{figure}[t]
\includegraphics[width=0.45 \textwidth]{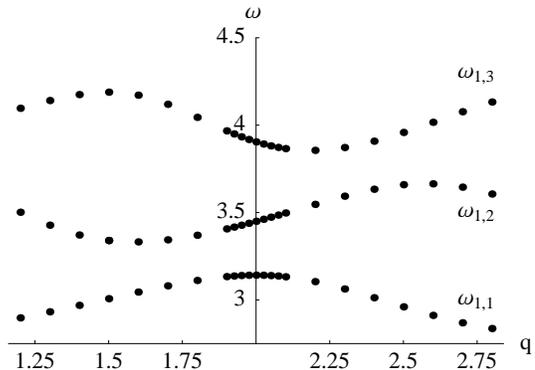}
\caption{\label{fig:spec4mcq} A portion of the optical frequency
spectrum for the two-membrane cavity. A closely spaced triplet of
frequencies $\omega_{1,i}(i=1,2,3)$ is shown, using a higher
resolution near the mode equilibrium point $q_0=2$. For $Q \neq 0$
the optical mode frequencies are similarly modulated in a
direction orthogonal to $q$. The cavity parameters used in the
plot are $R=0.5,L=1$.}
\end{figure}

Equation~(\ref{eq:trans}) can be solved numerically for a given
set of system parameters. Figure ~\ref{fig:spec4mcq} shows a
triplet of optical modes $\omega_{n,i}(q, Q)$ in the vicinity of
the equilibrium point of relative motion $q_0=2$ and for $Q=0$.
The optical frequencies are strongly modulated along $q$, a
situation familiar from the case of a single membrane, $N=1$.
These modulations arise from the avoided crossings associated with
the lifting of the degeneracy of the frequency triplet due to the
coupling of the sub-cavities \cite{mishpm7}. Similar periodic
modulations appear in $\omega_{n,i}(q, Q)$ for fixed $q$ and
$Q$ varying.

Considerable insight can be gained from approximate analytical
solutions of Eq.~(\ref{eq:trans}) \cite{mishpm7}. Perturbation
theory in the small parameters $(q-q_0,Q-Q_0) \ll 2\pi c/\omega_n$
yields for example the triplet $(i=1,2,3)$ of optical frequencies
in the vicinity of any $q_0$ and $Q_0$
\begin{eqnarray}
 \label{eq:freq}
\omega_{n,i}(q,Q)& = &\Delta_{n,i}+B_{n,i}(q-q_0)+B_{n,i}'(Q-Q_0)\nonumber \\
&&+M_{n,i} (q-q_0)^{2}+M_{n,i}'(Q-Q_0)^{2}  \\
&&+P_{n,i}(q-q_0)(Q-Q_0)+\ldots.\nonumber \\ \nonumber
\end{eqnarray}

The various terms in this equation fully describe the 
basic optomechanical properties of the system. Specifically,
$\Delta_{n,i}$ are the frequency shifts due to the sub-resonators
coupling in the absence of membrane motion. $B_{n,i}$ and
$B_{n,i}'$ determine the strength of the linear optomechanical
couplings associated with the ``breathing" and COM modes of
motion, respectively, producing a back-action of the mirror 
motion on the light field that can be exploited in mirror 
cooling. $M_{n,i}$ and $M_{i}'$ govern the quadratic couplings 
and can lead to quantum non-demolition (QND) energy measurements 
\cite{thompson2008}). Finally $P_{n,i}$ quantifies the coupling
between the relative and COM modes and is responsible for normal 
mode decoherence as well as down-conversion \cite{james2003}. We 
discuss later in this Letter circumstances under which that 
coupling can be cancelled.

Consider for concreteness the case $q_0=2L, Q_0=0$. In this case
we have
\begin{eqnarray}
 \label{eq:shifts}
\Delta_{n,1}&=&\frac{n\pi c}{2L},\nonumber \\
\Delta_{n,2}&=&\frac{c}{2L}\left[n\pi
+\sin^{-1}\left(\frac{\sqrt{3+\sin^{2} \theta}}{2}\right)-\theta \right], \\
\Delta_{n,3}&=&\frac{c}{2L}\left[(n+1)\pi
-\sin^{-1}\left(\frac{\sqrt{3+\sin^{2} \theta}}{2}\right)-\theta \right],\nonumber \\
\nonumber
\end{eqnarray}
and
\begin{eqnarray}
\label{eq:linocq}
&&B_{n,i}=\nonumber \\
&&\frac{\xi_{n,i} \sin (\theta_{n,i} ) \sin (2 \theta )}
{3 \cos (\theta_{n,i}) \sin^{2}(\theta)+3 \cos (2\theta+3 \theta_{n,i})+
\sin (2 \theta ) \sin (\theta_{n,i}) }, \nonumber\\
\end{eqnarray}
where $\theta_{n,i}=2 \Delta_{n,i}L/c$ and
$\xi_{n,i}=\theta_{n,i}/(2 L)^{2}$.

In that case, both $B_{n,2}$ and $B_{n,3}$, which are the slopes
of the two upper curves in Fig.~\ref{fig:spec4mcq}, are different
from zero. The linear spatial dependence of the optical
frequencies on $q$ can therefore be used for back-action cooling
and trapping of the ``breathing" mode of the pair of membranes. In
contrast, $B_{n,1}=0$, and the corresponding resonator mode cannot
be used for that purpose. However it can be used for a
non-demolition measurement of the $q$ mode, as it provides a
purely quadratic coupling
\begin{equation}
 \label{eq:Mn}
M_{n,1}=-\frac{1}{6}\tau \xi_{n,1}^{2} \tan \theta \neq 0,
\end{equation}
where $\tau=4L/c$ is the round trip time for each sub-cavity.
Hence both linear and quadratic optomechanical couplings are
accessible at the same point $q_0=2$ using modes from the same
frequency triplet. This is unlike the situation for a single
membrane, see Refs. \cite{thompson2008,mishpm2}.

The analysis for the mode $Q$ is similar to the case of a single
membrane since both describe the effects of COM motion
\cite{mishpm2}. The choice of $Q_0=0$ implies the absence of the
terms linear in $Q$ in the expansion of Eq.~(\ref{eq:freq}), a
direct consequence of the fact that Eq.~(\ref{eq:trans}) is
symmetric about $Q_0=0.$ Hence $B_{n,i}'=0$. The quadratic
coupling is however non-zero; for reasons of space we will
provide an expression for it elsewhere \cite{mishpm7}.
On the other hand, when the membranes are positioned so that $Q_0
\neq 0$ then non-vanishing terms linear in $Q$ arise in
Eq.~(\ref{eq:freq}) and optomechanical center-of-mass cooling
becomes possible. Further analysis of this case will be presented
elsewhere \cite{mishpm7}.

We finally turn to the last term in Eq.~(\ref{eq:freq}). As
already mentioned, it describes the coupling between the breathing
and COM modes of motion of the membranes. In perturbation theory
this bilinear term is smaller than the linear terms, hence it does
not effect significantly the back-action properties used in
cooling.  Furthermore, for $Q_0=0$ we must have that $P_{n,i}=0$
since Eq.~(\ref{eq:trans}) is even in $Q$ and hence the expansion
cannot contain terms that are odd in $Q$. To lowest order, the
coupling between these modes of motion is then proportional to
$(q-q_o)Q^{2}$ and is negligible. In that case, the breathing and
COM modes remain to an excellent approximation the `true' normal
modes of the full two-membrane optomechanical problem. This is a 
central result of this paper. In physical terms, this implies 
that with an appropriate combination of optical frequencies, it 
is possible to independently cool, trap and characterize the 
breathing and COM modes of the 2-membrane array. 

We now discuss in broad terms the extension of these
considerations to an array of $N>2$ membranes. The addition of a
membrane positioned at a node of the intracavity field does not
change the optical finesse measurably ; if the mirror is placed
away from a frequency extremum, the finesse is multiplied by
$R\sim 0.99$ \cite{thompson2008,mishpm2}. Thus a scaling up to
$N\sim 10$ membranes should be practical.

The boundary conditions of the optical resonator yield a
trigonometric equation similar to Eq.~(\ref{eq:trans})
independently of the number $N$ of membranes. It is always
possible to express this equation in terms of the $N$ normal modes
$Q_{j}$ of the array, since they are linear combinations of the
individual oscillator coordinates $q_{j}$. Generally the resulting
optical spectrum will be periodic in every mechanical mode,
guaranteeing the existence of ranges where the dependence of the
$N+1$ frequencies $\omega_{n,i}(\{Q_j\}), i=1, \dots N+1,$
exhibits extrema, and other regions of linear dependence. This in
turn implies the possibility of cooling, trapping and performing
QND measurements of the energy of all collective modes. The
optomechanical parameters coupling any optical mode to any
mechanical mode can be extracted either analytically, from
expressions such as Eqs.~(\ref{eq:freq}-\ref{eq:Mn}), or
numerically. The cooling and trapping effects due to back-action ,
being linear in the $Q_j$, are always independent of the
coupling between the phonon modes.

Most importantly, we expect the optical spectrum to be symmetric
in every symmetric normal mode of the array. We have shown this
explicitly for the COM mode in the cases $N=1$ \cite{mishpm2} and
$N=2$ (this work). There are $N/2$ symmetric modes for $N$ even,
and $N/2+1$ for $N$ odd \cite{james1998}. For $N=3$ for example,
with individual oscillator coordinates $q_{j},(j=1,2,3)$ we expect
the optical spectrum to be symmetric in the COM mode
$Q_{1}=(q_1+q_2+q_3)/3$ as well as in the `scissors' mode
$Q_{2}=(q_1-2q_2+q_3)/6$, but not in the stretch mode
$Q_{3}=(q_3-q_1)/2$ \cite{mishpm7}. However the amount of symmetry
available is adequate to cancel all mode-mode coupling terms by
appropriate positioning of the membranes. For $N \geq 4$, some
mode coupling terms will have to be included in the description of
the optomechanics. For $N$ even, for example, ${N/2 \choose
2}=N(N-2)/8$ coupling terms out of a total possible ${N \choose
2}=N(N-1)/2$ will remain in the expression for each frequency.

The full behavior of the system may be modelled by a Hamiltonian
analogous to Eq.~(\ref{eq:Ham}), combined with a quantum treatment
of the noise associated with the input fields and with
dissipation, see e.g. Ref. \cite{girvin2007}. Experimentally the
noise spectrum of any collective modes can be obtained in
principle by modulating the light field at the specific mode
frequency \cite{gigan2006}.

Our description of the membrane array is reminiscent of a chain of
trapped ions \cite{zoller1995,wineland1998,james1998,james2003}, 
with however significant differences. First, in the present 
case the coupling between mirrors is due to radiation pressure 
rather than the Coulomb interaction, and is therefore switchable. 
As such, the situation is perhaps more akin to the case of 
dipolar molecules, where the coupling is due to the dipole-dipole 
interaction and can also be tuned. Also, membranes and mirrors do 
not possess internal degrees of freedom, at least at the level of 
the present description. Still, the analogy to the physics of 
trapped ions or dipolar molecules on lattices raises the possibility 
of using multi-membrane systems as information processing devices. 
Each membrane can be addressed individually through its suspension, 
and additional degrees of freedom such as rotation could possibly be
added and exploited \cite{mishpmrot}. Issues of noise and
decoherence will of course be extremely challenging, but some work
along these lines is already being carried out in efforts to cool
such systems to their vibrational ground state
\cite{kippenberg2007,girvin2007}. In addition, as is the case for
trapped ions the independent control of the collective normal
modes of the array will become technically increasingly
challenging with increasing $N$.

In conclusion we have considered the extension of cavity
optomechanics to multi-membrane systems, and found that the
normal modes of the system can cooled, trapped and measured. The
similarity of this system to chains of trapped ions or dipoles 
leads us to conjecture that it may find possible future 
application as an information-processing device operated in 
the quantum regime.

This work is supported in part by the US Office of Naval Research,
by the National Science Foundation, and by the US Army Research
Office. We would like to thank H. Uys and O. Dutta for stimulating
discussions.



\begin{thebibliography}{34}
\expandafter\ifx\csname natexlab\endcsname\relax\def\natexlab#1{#1}\fi
\expandafter\ifx\csname bibnamefont\endcsname\relax
  \def\bibnamefont#1{#1}\fi
\expandafter\ifx\csname bibfnamefont\endcsname\relax
  \def\bibfnamefont#1{#1}\fi
\expandafter\ifx\csname citenamefont\endcsname\relax
  \def\citenamefont#1{#1}\fi
\expandafter\ifx\csname url\endcsname\relax
  \def\url#1{\texttt{#1}}\fi
\expandafter\ifx\csname urlprefix\endcsname\relax\def\urlprefix{URL }\fi
\providecommand{\bibinfo}[2]{#2}
\providecommand{\eprint}[2][]{\url{#2}}

\bibitem[{coh()}]{cohadon1999}
\bibinfo{note}{P. F. Cohadon, A. Heidmann, and M. Pinard, Phys. Rev. Lett.
  \textbf{83}, 3174 (1999).}

\bibitem[{kle()}]{kleckner2006}
\bibinfo{note}{D. Kleckner and D. Bouwmeester, Nature \textbf{444}, 75 (2006).}

\bibitem[{gig()}]{gigan2006}
\bibinfo{note}{S. Gigan  \textit{et. al}, Nature \textbf{444}, 67 (2006).}

\bibitem[{arc()}]{arcizet2006}
\bibinfo{note}{O. Arcizet, P. -F. Cohadon, T. Briant, M. Pinard, and A.
  Heidmann, Nature \textbf{444}, 71 (2006).}

\bibitem[{sch()}]{schliesser2006}
\bibinfo{note}{A. Schliesser, P. Del'Haye, N. Nooshi, K. J. Vahala, and T. J.
  Kippenberg, Phys. Rev. Lett. \textbf{97}, 243905 (2006).}

\bibitem[{cor()}]{corbitt2007}
\bibinfo{note}{T. Corbitt \textit{et. al}, Phys.
  Rev. Lett \textbf{98}, 150802 (2007).}

\bibitem[{mar()}]{marshall2003}
\bibinfo{note}{W. Marshall, C. Simon, R. Penrose, and D. Bouwmeester, Phys.
  Rev. Lett. \textbf{91}, 130401 (2003).}

\bibitem[{rou()}]{roukes2006}
\bibinfo{note}{Y. T. Yang \textit{et. al}, Nano Lett. \textbf{6}, 583 (2006).}

\bibitem[{cou()}]{courty2003}
\bibinfo{note}{J.-M. Courty, A. Heidmann, and M. Pinard, Phys. Rev. Lett.
  \textbf{90}, 083601 (2003).}

\bibitem[{han()}]{hansch2007}
\bibinfo{note}{P. Treutlein, D. Hunger, S. Camerer, 
T. W. Hansch, and J. Reichel, Phys. Rev. Lett. \textbf{99}, 140403 (2007).}

\bibitem[{man()}]{mancini2003}
\bibinfo{note}{S. Mancini, D. Vitali and P. Tombesi, Phys.
  Rev. Lett. \textbf{90}, 137901 (2003).}

\bibitem[{\citenamefont{Braunstein et~al.}(2000)
\citenamefont{Braunstein, and H.-K. Lo}}]{Braunsteinbook2}
  \bibinfo{author}{\bibfnamefont{S.~L.} \bibnamefont{Braunstein}},
  \bibnamefont{and} \bibinfo{author}{\bibfnamefont{H.~-K.}
  \bibnamefont{Lo}},
\emph{\bibinfo{title}{Scalable Quantum Computers: paving the way to realization}}
  (\bibinfo{publisher}{Wiley},
  \bibinfo{address}{Weinheim}, \bibinfo{year}{2000}).

\bibitem[{zol()}]{zoller1995}
\bibinfo{note}{J. I. Cirac and P. Zoller, Phys. Rev. Lett. \textbf{74}, 4091 (1995).}

\bibitem[{win()}]{wineland1998}
\bibinfo{note}{D. Wineland \textit{et. al}, J.
  Res. Natl. Inst. Stand. Tech., \textbf{103}, 259 (1998).}

\bibitem[{dem()}]{demille2002}
\bibinfo{note}{D. DeMille, Phys. Rev. Lett. \textbf{88}, 067901 (2002).}

\bibitem[{tho()}]{thompson2008}
\bibinfo{note}{J. D. Thompson \textit{et. al}, Nature
  \textbf{452}, 72 (2008).}

\bibitem[{kip()}]{kippenberg2007}
\bibinfo{note}{I. Wilson-Rae, N. Nooshi, W. Zwerger and T. J. Kippenberg,
Phys. Rev. Lett. \textbf{99}, 093901 (2007).}

\bibitem[{gir()}]{girvin2007}
\bibinfo{note}{F. Marquardt, J. P. Chen, A. A. Clerk and S. M. Girvin,
Phys. Rev. Lett. \textbf{99}, 093902 (2007).}

\bibitem[{ane()}]{anetsberger2007}
\bibinfo{note}{A. Schliesser, R. Riviere, G. Anetsberger, O. Arcizet and T. J. Kippenberg,
arXiv:0712.1618v1 [physics.optics]}

\bibitem[{bra()}]{braginsky1980}
\bibinfo{note}{V. B. Braginsky, Y. I. Vorontsov and K. S. Thorne,
Science \textbf{209}, 547 (1980).}

\bibitem[{mis()}]{mishpm2}
\bibinfo{note}{M. Bhattacharya, H. Uys and P. Meystre, Phys. Rev. A
\textbf{77}, 033819 (2008).}

\bibitem[{mis()}]{mishpm7}
\bibinfo{note}{M. Bhattacharya and P. Meystre, in preparation.}

\bibitem[{jam()}]{james2003}
\bibinfo{note}{D. F. V. James, App. Phys. B \textbf{76}, 199 (2003).}

\bibitem[{jam()}]{james1998}
\bibinfo{note}{D. F. V. James, App. Phys. B
\textbf{66}, 181 (1998).}

\bibitem[{mis()}]{mishpmrot}
\bibinfo{note}{M. Bhattacharya and P. Meystre,
Phys. Rev. Lett. \textbf{99}, 153603 (2007).}

\end{thebibliography}
\end{document}